\documentclass[preprint,amsmath,amssymb,superscriptaddress,unsortedaddress]{revtex4}
\usepackage{color}
\usepackage[normalem]{ulem}
\usepackage{newfloat}
\usepackage{endnotes}
\usepackage{amsmath}
\usepackage{graphicx}
\usepackage{setspace}
\makeatletter
\let\saved@includegraphics\includegraphics
\AtBeginDocument{\let\includegraphics\saved@includegraphics}
\renewenvironment*{figure}{\@float{figure}}{\end@float}
\makeatother
\begin{document}
%%%%%%%%%%
%%	Title
%%%%%%%%%%
\title{Multipole polaron in the devil's staircase of CeSb}
%%%%%%%%%%
%%	Authors
%%%%%%%%%%
\author{Y.~Arai}
\affiliation{Institute for Solid State Physics, The University of Tokyo, Kashiwa, Japan}
\author{Kenta~Kuroda\footnote[2]{Present address: Graduate School of Advanced Science and Engineering, Hiroshima University, Higashihiroshima, Japan}\footnote[3]{Corresponding author: kuroken224@hirsohima-u.ac.jp}}
\affiliation{Institute for Solid State Physics, The University of Tokyo, Kashiwa, Japan}
\author{T.~Nomoto} 
\affiliation{Department of Applied Physics, The University of Tokyo, Tokyo, Japan}
\author{Z.~H.~Tin} 
\affiliation{Department of Physics, Osaka University, Toyonaka, Japan}
\author{S.~Sakuragi} 
\affiliation{Institute for Solid State Physics, The University of Tokyo, Kashiwa, Japan}
\author{C.~Bareille} 
\affiliation{Institute for Solid State Physics, The University of Tokyo, Kashiwa, Japan}
\author{S.~Akebi} 
\affiliation{Institute for Solid State Physics, The University of Tokyo, Kashiwa, Japan}
\author{K.~Kurokawa} 
\affiliation{Institute for Solid State Physics, The University of Tokyo, Kashiwa, Japan}
\author{Y.~Kinoshita} 
\affiliation{Institute for Solid State Physics, The University of Tokyo, Kashiwa, Japan}
\author{W.-L.~Zhang}
\affiliation{Institute for Solid State Physics, The University of Tokyo, Kashiwa, Japan}
\affiliation{Department of Engineering and Applied Sciences, Sophia University, Tokyo, Japan}
\author{S.~Shin} 
\affiliation{Institute for Solid State Physics, The University of Tokyo, Kashiwa, Japan}
\affiliation{Office of University Professor, The University of Tokyo, Kashiwa, Japan}
\author{M.~Tokunaga} 
\affiliation{Institute for Solid State Physics, The University of Tokyo, Kashiwa, Japan}
\affiliation{Trans-scale Quantum Science Institute, The University of Tokyo, Tokyo, Japan}
\author{H.~Kitazawa} 
\affiliation{National Institute for Materials Science, Tsukuba, Japan}
\author{Y.~Haga} 
\affiliation{Advanced Science Research Center, Japan Atomic Energy Agency, Tokai, Japan}
\author{H.~S.~Suzuki} 
\affiliation{Institute for Solid State Physics, The University of Tokyo, Kashiwa, Japan}
\author{S.~Miyasaka} 
\affiliation{Department of Physics, Osaka University, Toyonaka, Japan}
\author{S.~Tajima} 
\affiliation{Department of Physics, Osaka University, Toyonaka, Japan}
\author{K.~Iwasa}
\affiliation{Frontier Research Center for Applied Atomic Sciences and Institute of Quantum Beam Science, Ibaraki University, Tokai, Japan}
\author{R.~Arita} 
\affiliation{Department of Applied Physics, The University of Tokyo, Tokyo, Japan}
\affiliation{RIKEN Center for Emergent Matter Science (CEMS), Wako, Japan}
\author{Takeshi~Kondo} 
\affiliation{Institute for Solid State Physics, The University of Tokyo, Kashiwa, Japan}
\affiliation{Trans-scale Quantum Science Institute, The University of Tokyo, Tokyo, Japan}
\maketitle
%%%%%%%%%%
%%	Abstract
%%%%%%%%%%
\textbf{
Rare-earth intermetallic compounds exhibit rich phenomena induced by the interplay between localized $\textit{\textbf{f}}$ orbitals and conduction electrons.
However, since the energy scale of the crystal-electric-field splitting is only a few millielectronvolts, the nature of the mobile electrons accompanied by collective crystal-electric-field excitations has not been unveiled.
Here, we examine the low-energy electronic structures of CeSb through the anomalous magnetostructural transitions below the N{\'{e}}el temperature, $\mathbf{\sim\!17\;\mathrm{\mathbf{K}}}$, termed the `devil's staircase', using laser angle-resolved photoemission, Raman and neutron scattering spectroscopies.
We report another type of electron-boson coupling between mobile electrons and quadrupole crystal-electric-field excitations of the 4$\textit{\textbf{f}}$ orbitals, which renormalizes the Sb 5$\textit{\textbf{p}}$ band prominently, yielding a kink at a very low energy ($\mathbf{\sim\!7\;\mathrm{\mathbf{meV}}}$).
This coupling strength is strong and exhibits anomalous step-like enhancement during the devil's staircase transition, unveiling a new type of quasiparticle, named the `multipole polaron', comprising a mobile electron dressed with a cloud of the quadrupole crystal-electric-field polarization.
}

%%%%%%%%%%
%Main text
%%%%%%%%%%
Interesting properties of condensed matter can emerge from strong many-body interactions, and one of the relevant interactions is an electron-boson coupling that leads to a quasiparticle (QP) of an electron combined with a local bosonic field.
The investigation of such a QP is important since its dynamical behaviour determines fundamental physical quantities like electrical conductivity and often plays roles in electronic instabilities leading to strongly correlated phenomena such as high-temperature superconductivity~\cite{Reznik_nature2006} and colossal magnetoresistance~\cite{Teresa_nature1997}.
One of the most famous QPs is the polaron---an electron dressed with a local polarization (phonon) in the background lattice \cite{Millis_Nature1998}, but such a QP state can be realized not only by electron-phonon coupling but also by coupling to other bosonic modes \cite{Devreese_polaron2003}.
The best quantity for exploring the electron-boson couplings is the renormalization of the QP band with a kink that is directly accessible by angle-resolved photoemission spectroscopy (ARPES) \cite{Damascelli_rmp2003}.
However, despite a number of investigations, only three types of electron-boson coupling have been experimentally identified so far; electron-phonon (or polaron) \cite{Valla_prl1999,Lanzara_Nature2001}, electron-magnon (or magnetic polaron) \cite{Schafer_prl2004,Dahm_NaturePhys2009}, and electron-plasmon (or plasmaron) \cite{Bostwick_science2010} couplings.

Here we found a new type of electron-boson coupling by investigating a rare-earth intermetallic compound CeSb, which leads to a unique QP state named the `multipole polaron'.
This new coupling mediates between the itinerant electrons and the bosons constituted by the crystal-electric-field (CEF) excitations of the localized 4$f$ states.
As evidence for it, our high-resolution ARPES combined with a laser source reveals a low-energy kink of the Sb 5$p$ band at $\sim\!7\;\rm{meV}$ and  displays the mass enhancement quantitatively defined by the coupling constant ($\lambda$) of the many-body interactions.
We show that this low-energy feature corresponds to a quadrupole CEF excitation ($J_z\!=\!\pm2$) in the $J$ multiplets of the 4$f$ orbitals, possessing an electronic-magnetic unified degree of freedom, the `multipole'.
Notably, the discovered multipole polaron shows an extremely strong coupling ($\lambda\!\approx\!4$) far beyond the limit for the strong electron-phonon coupling systems ($\lambda\!\sim\!2$) and exhibits step-like enhancement in $\lambda$ with the anomalous phase transition of the ordered 4$f$ states called the `devil's staircase'.

\begin{figure}
	\centering
	\includegraphics[width=0.75\textwidth]{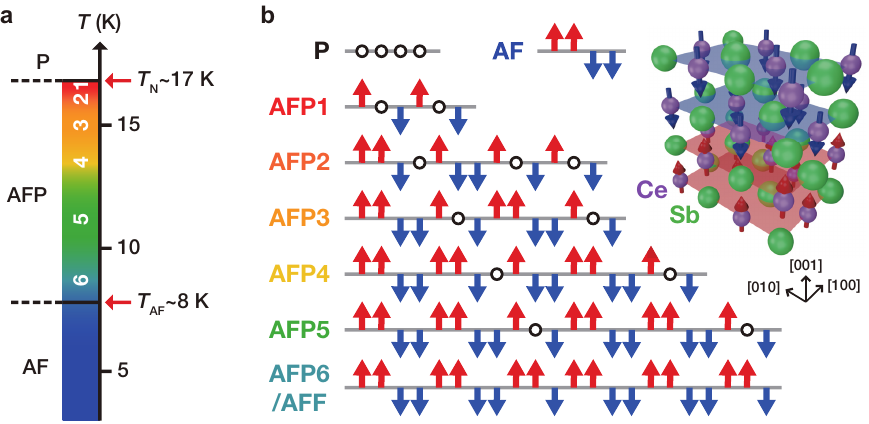}
	\caption{\label{fig1}
	\textbf{Magnetostructures established in the devil's staircase of CeSb}.
	(\textbf{a}) Temperature scale of the devil's staircase transition at zero field.
	(\textbf{b}) Schematics of magnetostructures \cite{Mignod_prb1977,Fischer_jpc1978} that comprise stackings of ferromagnetic (arrows) and paramagnetic (circles) planes along the crystallography axis [001].
	The inset shows the magnetic texture for the magnetic ground state (the AF phase).
	The structure of the AFP6 phase has not yet been determined.
	The transition from the AFP5 phase to the AFP6 phase occurs without a clear change of the $\bf{q}$ vector but with a large entropy variation \cite{Mignod_jpc1980}.
	This transition was interpreted as an elimination of the paramagnetic planes, and instead of the AFP6 phase, the AFF structure is considered (namely the AFP6/AFF phase).
	The ordered Ce ions have large moments of $\sim\!2\;\mu_{\rm{B}}$ ($\mu_{\rm{B}}$, the Bohr magneton) along the [001] axis, which is consistent with an almost fully polarized $f\Gamma_{80}$ state with $J_z\!=\!5/2$  (see Fig.~\ref{fig3}\textbf{a}) \cite{Takahashi_jpc1985,Mignod_jmmm1985}.
	On the other hand, the other Ce ions in the paramagnetic planes leave the $f\Gamma_{7}$ ground state corresponding to the $\sim\!0.7\;\mu_{\rm{B}}$ moments.
These competing CEF ground states imply a spatial modulation of the $p$-$f$ mixing \cite{Iwasa_prl2002} (details in text).
}
\end{figure}

CeSb has a simple NaCl crystal structure but presents rather complex phase transitions, the so-called devil's staircase \cite{Bak_PhysToday1986}, in which many phases with different long-periodic magnetostructures sequentially appear with temperatures below the N{\'{e}}el temperature ($T_{\rm{N}}$) of $\sim\!17\;\rm{K}$ (Fig.~\ref{fig1})~\cite{Mignod_prb1977, Fischer_jpc1978}.
They comprise stackings of the Ising-like ferromagnetic (001) planes and paramagnetic planes.
The antiferromagnetic (AF) structure with a double-layer modulation is established below $\sim\!8\;\rm{K}$ ($T_{\rm{AF}}$) as the magnetic ground state, while six antiferro-paramagnetic (AFP) phases, where the double-layer AF modulation is periodically locked by the paramagnetic (P) layer, appear between $T_{\rm{AF}}$ and $T_{\rm{N}}$.
The transition from the AFP5 to AFP6 phase occurs with a large variation of the entropy but without a clear change of the magnetic propagating vector ($\bf{q}$)~\cite{Mignod_jpc1980}. 
Thus, instead of the AFP6 phase, the antiferro-ferromagnetic (AFF) structure without the paramagnetic planes was previously considered (we call it the AFP6/AFF phase hereafter; Fig~\ref{fig1}\textbf{b}).
Indeed, the AFF phase is appropriate as a precursor of the AF ground state, since other AFF phases with different $\bf{q}$ show up as the lowest-temperature phase under a magnetic field~\cite{Mignod_prb1977}.
Accompanying these magnetostructural transitions in the devil's staircase, the established 4$f$ order induces a tetragonal lattice contraction \cite{hulliger_jltp1975,Iwasa_prl2002}, dramatic changes of electron transports \cite{Mori_jap1991,Ye_prb2018,Xu_nacom2019}, electronic reconstructions \cite{Kumigashira_prb1997,Takayama_jpsj2009,Jangeaat_scienceAd2019,kuroda2020NatureComm,Nishi_prb2005} and a pseudo-gap-like anomaly in the momentum-resolved spectral function \cite{kuroda2020NatureComm}.
All of these facts suggest the importance of the interplay between the localized 4$f$ states and the itinerant electrons \cite{Takahashi_jpc1985, Kasuya_lowCarrier1995} for understanding the anomalous transitions.
However, revealing the underlying many-body interaction going through the devil's staircase has remained out of reach, to the best of our knowledge.

\begin{figure*}
	\centering
	\includegraphics[width=\textwidth]{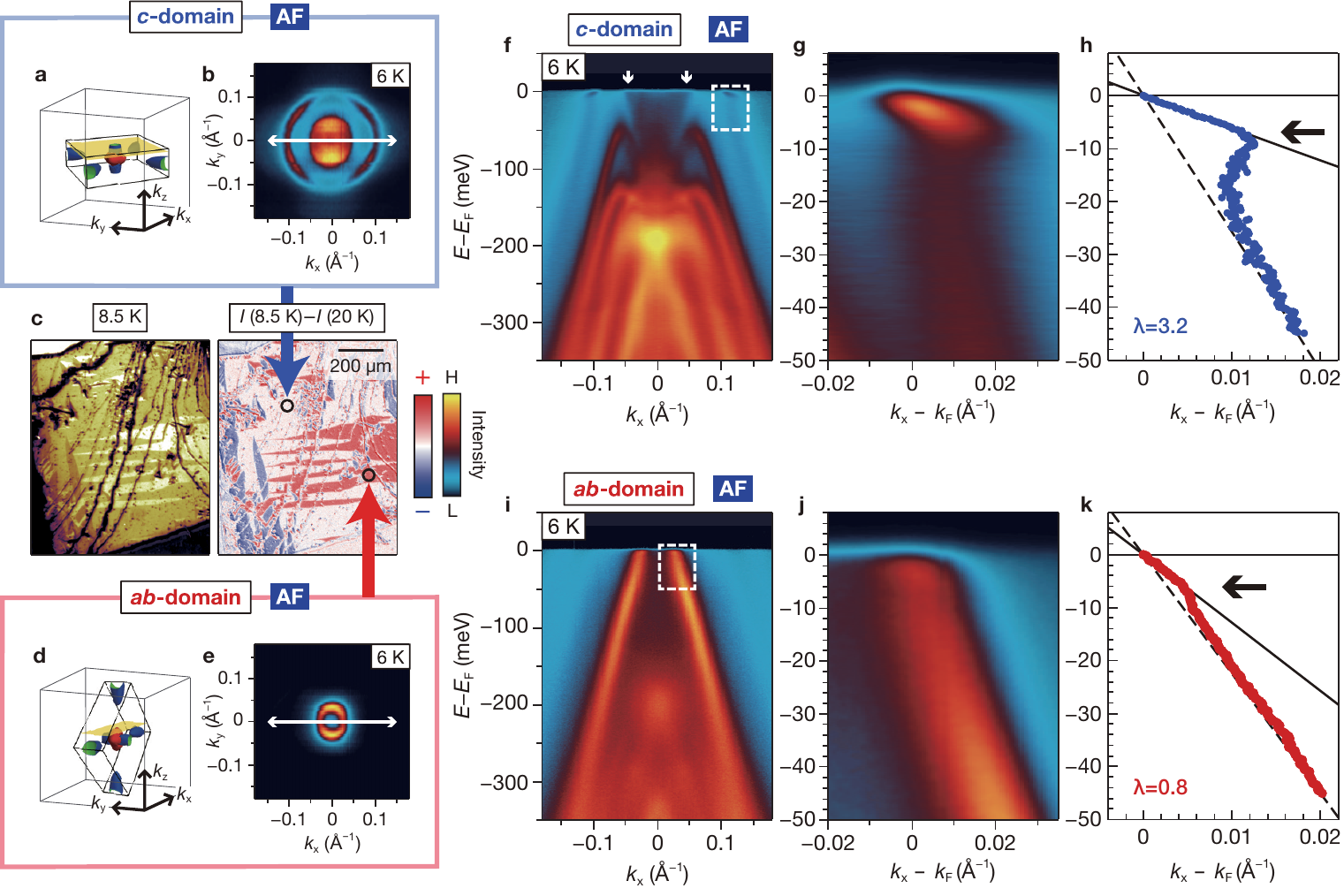}
	\caption{
\label{fig2}
	\textbf{Electron-boson coupling of the Sb 5\textit{\textbf{p}} hole band.}
	(\textbf{a}-\textbf{d}) Calculated Fermi surfaces in the folded Brillouin zone according to the AF modulation for different magnetic domains, namely the $c$ and $ab$ domains (\textbf{a} and \textbf{d}) \cite{kuroda2020NatureComm}.
	$k_{x}$, $k_{y}$ and $k_{z}$ are electron wavevectors along [100], [010] and [001] axes.
	These domains are spatially mapped by polarized microscope images of the cleaved (001) surface in \textbf{c}.
	The left image is obtained below $T_{\rm{N}}$ ($8.5\;\rm{K}$), and the right image represents the intensity ($I$) difference below and above $T_{\rm{N}}$.
	Our laser ARPES with a well-focused laser spot (black circles) distinctively observes the bands in $c$ domain (white area) and $ab$ domain (red and blue areas).
	The yellow planes in \textbf{a} and \textbf{d} represent the $k_x$-$k_y$ plane at $k_z\!\approx\!0.2\;\rm{\AA}^{-1}$ detected by our laser ARPES \cite{kuroda2020NatureComm}.
	Panels \textbf{b} and \textbf{e} show the experimental Fermi surface maps at $T\!=\!6\;\rm{K}$ (AF phase).
	H, high; L, low.
	(\textbf{f}-\textbf{k}) Panels \textbf{f} and \textbf{i} show the band maps cut along the white arrows in \textbf{b} and \textbf{e}.
	$k_{\rm{F}}$, Fermi wave number.
	Only in the $c$ domain is the folded electron band that is derived from Ce 5$d$ states observed, as denoted by the arrows in \textbf{f}.
	Panels \textbf{g} and \textbf{j} show the magnified images for the prominent kink in the band dispersion within the dashed rectangles shown in \textbf{f} and \textbf{i}. 
	Panels \textbf{h} and \textbf{k} show the peak plots to trace the renormalized QP bands (solid lines) and the bare bands (dashed line, Supplementary Note 1).
	Arrows highlight the bosonic-mode energy.
}
\end{figure*}

We present laser ARPES maps at $6\;\rm{K}$ (the AF phase) in Fig.~\ref{fig2}\textbf{f}, \textbf{i}.
In the same manner as in our previous report \cite{kuroda2020NatureComm}, we distinguish two domains with an ordered 4$f$ moment either along the perpendicular to the cleaved (001) plane ($c$ domain; Fig. \ref{fig2}\textbf{a}, \textbf{b}) or on the plane ($ab$ domain; Fig. \ref{fig2}\textbf{d}, \textbf{e}), which can be identified in our polarized microscope images (Fig.~\ref{fig2}\textbf{c}).
Below $T_{\rm{N}}$, since the electronic symmetry reduces to tetragonal due to back-folding by the magnetic 4$f$ order (Fig. \ref{fig2}\textbf{a}, \textbf{d}), the electronic anisotropy appears as contrastive dispersions \cite{kuroda2020NatureComm}; only the hole band derived from the Sb 5$p$ state is observed in the $ab$ domain (Fig.~\ref{fig2}\textbf{i}), while the folded electron band derived from the Ce 5$d$ state is observed near the Fermi energy $E_{\rm{F}}$ (arrows in Fig.~\ref{fig2}\textbf{f}) together with the Sb 5$p$ hole band in the $c$ domain.

Thanks to the bulk sensitivity of our low-energy laser \cite{ShimojimaJPSJ2015} as well as the small 4$f$ photoionization cross-section \cite{YehADNDT1985}, our laser ARPES enables us to unambiguously reveal the many-body effect of the itinerant carriers.
A clear signature for the band renormalization of the Sb 5$p$ band is presented in Fig.~\ref{fig2}\textbf{g}, where we show the magnified image for the $c$ domain (rectangle in Fig.~\ref{fig2}\textbf{f}).
The data display a sharp QP peak in the band near $E_{\rm{F}}$, and its dispersion is traced by the momentum distribution curve analysis (Fig.~\ref{fig2}\textbf{h}).
The Fermi velocity is largely reduced with a prominent kink at energy $E\!-\!E_{\rm{F}}=\!-7\!\pm\!1\;\rm{meV}$, which directly manifests the presence of an electron-boson coupling.

Notably, the emergence of a pronounced kink at low energy, in general, requires a very strong electron-boson coupling; hence, one can immediately predict from the data in Fig.~\ref{fig2}\textbf{h} that the magnitude of $\lambda$ must be quite large.
The $\lambda$ value can be determined by the real part of the self-energy ($\Sigma^{\prime}(E)$) with ${\partial \Sigma^{\prime}(E)}{/}{\partial E}{|}_{E=E_{\rm{F}}}$ (Supplementary Note~1) \cite{Damascelli_rmp2003}. 
The prominent kink in CeSb unveiled by our high-quality data reflects $\lambda\!\approx\!3.2$ (which becomes greater, up to $\approx\!4.0$, in the AFP6/AFF phase, as we will show later), which is extremely large compared to the systems with strong electron-phonon coupling, where the coupling strength is only ${\lambda}\!\sim\!2$ at most \cite{Kawamura_prb2020,Lanzara_Nature2001}. 

The unusual feature in our findings is not only the large ${\lambda}$ but also its sharp variation with momentum.
This signature is evident in Fig. \ref{fig2}\textbf{j}, \textbf{k}, which shows the dispersion kink in the $ab$ domain and significantly small $\lambda$ ($\approx\!0.8$).
Since our laser ARPES detects these domains in different momentum regions (Fig. \ref{fig2}\textbf{a}, \textbf{d}), the contrastive $\lambda$ for $c$ and $ab$ domains should indicate the strong momentum dependence of the electron-boson coupling.

The nature of electron-boson couplings observed in ARPES is often a subject of great debate because several modes share a similar energy range within $20\mathchar`-200\;\rm{meV}$ \cite{Lanzara_Nature2001,Dahm_NaturePhys2009}.
In general, optical phonons are responsible for kinks in the QP band structures due to electron-phonon couplings.
However, the very low-energy scale of the kink in CeSb is well distinguished from its optical phonon reported at $\sim\!18\;\rm{meV}$ \cite{Tutuncu_jpcm2007} and excludes this scenario.
In addition, our calculation for the electron-phonon coupling of LaSb, which is a comparable material to CeSb without 4$f$ electrons, evaluates $\lambda$ to be only 0.13 (Supplementary Note~2).
Considering these results, it is apparent that the strong electron-boson coupling in CeSb discovered in Fig.~\ref{fig2} cannot be explained by the conventional interaction with phonon, and therefore, one has to consider another degree of freedom.

\begin{figure*}
	\centering
	\includegraphics[width=\textwidth]{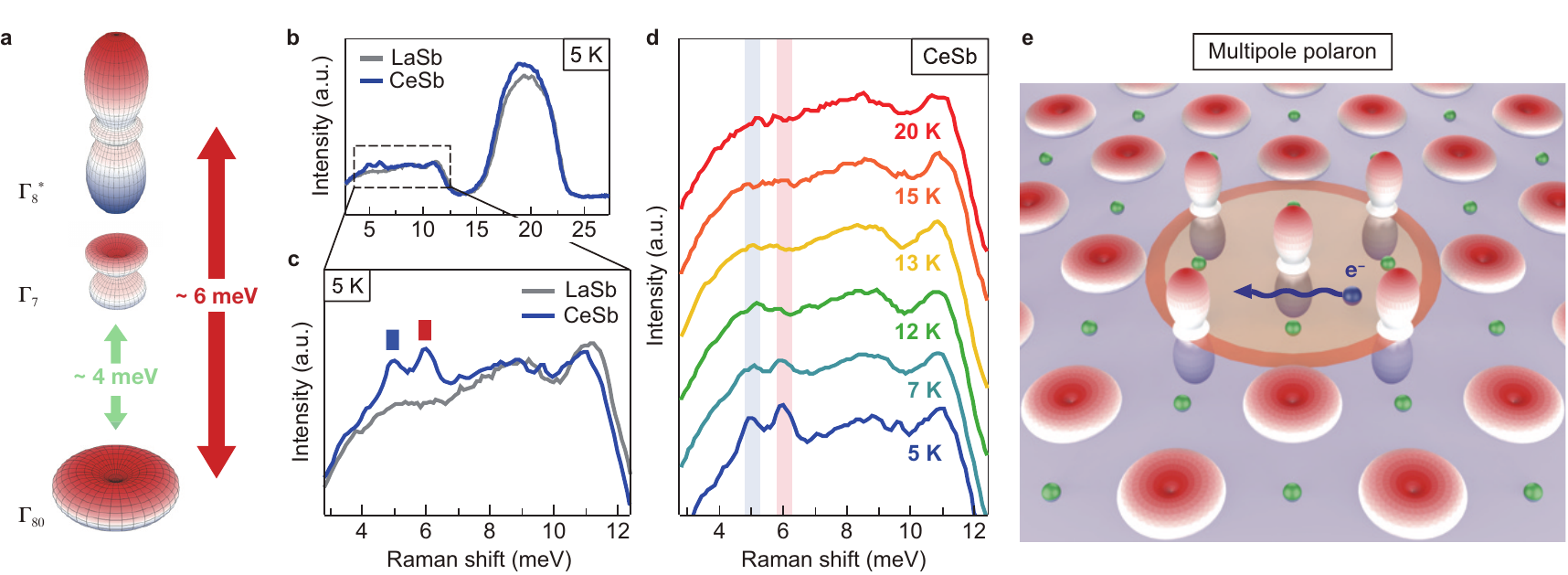}
	\caption{\label{fig3}
	\textbf{Raman spectra for CEF excitations and multipole polaron.}
	(\textbf{a}) CEF scheme of the localized 4$f$ states in the ordered phase.
	Above $T_{\rm{N}}$, the CEF ground state is the $f\Gamma_{7}$ doublet with the excited $f\Gamma_{8}$ quartet \cite{Mignod_jmmm1985}.
	Below $T_{\rm{N}}$, the $f\Gamma_{8}$ state splits into the almost fully polarized state with $J_z\!=\!5/2$ ($f\Gamma_{80}$) and the $f\Gamma_{8}$ with $J_z\!=\!1/2$ ($f\Gamma_{8}^{*}$), together with a small tetragonal lattice contraction along the 4$f$ moment \cite{hulliger_jltp1975,Iwasa_prl2002}.
	The schematics show the charge and magnetic charge densities of these wavefunctions of the $f\Gamma_{80}$, $f\Gamma_{7}$ and $f\Gamma_{8}^{*}$ orbits \cite{Kusunose_JPSJ2008}.
	The CEF splittings denoted by the red and green arrows are quantitatively determined by our Raman scattering and INS spectroscopies (also Supplementary Notes 3 and 4).
	(\textbf{b}) Raman spectra for LaSb (gray line) and CeSb (blue line) acquired at $T\!=\!5\;\rm{K}$.
 	(\textbf{c}) Enlarged spectra in the low-energy region denoted by the rectangle in \textbf{b}.
	Red and blue bars highlight peaks related to the quadrupole excitation of $f\Gamma_{80}$ to $f\Gamma_{8}^{*}$ (red arrow in \textbf{a}). 
	(\textbf{d}) Temperature dependence of the Raman spectra for $f\Gamma_{80}$ to $f\Gamma_{8}^{*}$ excitation.
	Red and blue bars highlight the same peaks highlighted in \textbf{c}.
	(\textbf{e}) Schematic of the multipole polaron, which is a composite QP propagating in the lattice of the lamellar $f\Gamma_{80}$ orbit (purple) as a free electron ($\rm{e}^{-}$) but largely dressed with the self-induced quadrupole CEF polarization of the longitudinal $f\Gamma_{8}^{*}$ orbit.
}
\end{figure*}

Figure \ref{fig3}\textbf{a} highlights the CEF splitting in the ordered 4$f$ states.
Across $T_{\rm{N}}$, an enhancement of the mixing between the Sb 5$p$ and 4$f$ states ($p$-$f$ mixing) \cite{Takahashi_jpc1985} splits the $f\Gamma_{8}$ quartet into the ground state with $J_z\!\approx\!5/2$ (hereafter $f\Gamma_{80}$) \cite{Mignod_jmmm1985} and the higher $f\Gamma_{8}$ states with $J_z\!\approx\!1/2$ (hereafter $f\Gamma_{8}^*$).
By previous inelastic neutron scattering (INS) spectroscopy, the dipole excitation ($|{\Delta}J_z|\!=\!1$) of $f\Gamma_{80}$ to $f\Gamma_{7}$ at $3\mathchar`-4\;\rm{meV}$ was observed as magnetic excitation~\cite{Halg_prb1986} (green arrow in Fig.~\ref{fig3}\textbf{a}).
However, we distinctly rule out this excitation for the electron-boson coupling observed in Fig.~\ref{fig2}, since its energy scale is too small compared with the kink energy of $\sim\!7\;\rm{meV}$.
To date, mainly INS spectroscopy was applied to examine the CEF splitting.
Therefore, we used high-resolution laser Raman scattering on CeSb and reveal that the quadrupole CEF excitation is responsible for the electron-boson coupling, as demonstrated in Fig.~\ref{fig3} (also Supplementary Notes 3 and 4).

The main results are presented in Fig. \ref{fig3}\textbf{b}, \textbf{c}, where we compare the Raman spectra of CeSb (blue) and LaSb (gray) obtained at $5\;\rm{K}$.
As the two compounds are different only in the presence/absence of the 4$f$ electron, their comparison allows us to disentangle signals related directly to the CEF excitations.
In Fig.~\ref{fig3}\textbf{b}, two common features exist, which therefore both reflect phonons \cite{Tutuncu_jpcm2007}: the broad peaks around $\sim\!18\;\rm{meV}$ and the continuum weights below $\sim\!12\;\rm{meV}$ (the details of these line shapes are clarified in Supplementary Note 3). 
Except for these trivial signals, one can clearly see that sharp low-energy peaks appear only in CeSb at $5\;\rm{meV}$ and $6\;\rm{meV}$ (Fig.~\ref{fig3}\textbf{c}).
We moreover find that these energies are robust against temperature (Fig.~\ref{fig3}\textbf{d}), but these peaks are suppressed with increasing temperature and eventually disappear around $T_{\rm{N}}$.
The lower CEF excitation of $f\Gamma_{80}$ to $f\Gamma_{7}$ is distinctively detected at $4\;\rm{meV}$ by our Raman scattering with the different light polarizations (Supplementary Note~3).
With these facts, we reach the only remaining assignment, that both of the two peaks at $5\;\rm{meV}$ and $6\;\rm{meV}$ derive from the quadrupole excitation of $f\Gamma_{80}$ to $f\Gamma_8^*$ (red arrow in Fig.~\ref{fig3}\textbf{a}).

The energy scale of the $f\Gamma_{80}$ to $f\Gamma_8^*$ excitation ($\sim\!5\mathchar`-6\;\rm{meV}$) shows good agreement with the kink energy ($7\!\pm\!1\;\rm{meV}$).
The higher excitation ($6\;\rm{meV}$) fits it better, but our self-energy analysis does not exclude the possible electron-boson coupling of both excitations (Supplementary Notes 1-3).
The very small deviation from the kink energy likely indicates the presence of the weak dispersion suggested by our INS spectroscopy (Supplementary Note~4).
We consider why double peaks are observed in the excitation of $f\Gamma_{80}$ to $f\Gamma_8^*$.
One possible interpretation is a vibronic coupling \cite{Thalmeier_elsiver1991} between the CEF and phonon excitations.
Such a hybrid excitation in CeSb is indeed suggested by INS spectroscopy \cite{iwasa2002apa} (Supplementary Notes 3 and 4).
This excitation can induce the doubling of excited states like bonding and antibonding states that may be detectable in Raman scattering \cite{Guntherodt_prl1983,Sethi_prl2019}.
Further measurements are necessary to validate this possible vibrionic coupling.

The quadrupole CEF excitation for $f\Gamma_{80}$ to $f\Gamma_8^*$ has multipole character and should be distinguished from the magnetic dipole excitation for $f\Gamma_{80}$ to $f\Gamma_7$ \cite{Halg_prb1986}.
Our observations of the prominent kink, therefore, reflect a coupling of the electron with the multipole CEF excitation.
Such a QP with the large mass enhancement can be described as a multipole polaron that propagates in a sea of lamellar $f\Gamma_{80}$ orbits but largely dressed with the self-induced quadrupole polarization of the longitudinal $f\Gamma_{8}^{*}$ orbits, as illustrated by Fig. \ref{fig3}\textbf{e}.
We use `polaron' for this itinerant QP characterized by the sharp dispersion kink, which should be distinguished from a Fr\"{o}hlich polaron observed as replica bands in ARPES spectra~\cite{Wang2016_NatureMat,Verdi_naturecom2017}.

One may wonder why such a many-body state is realized only through the excitation with $f\Gamma_{80}$ to $f\Gamma_8^*$ as opposed to $f\Gamma_{80}$ to $f\Gamma_{7}$.
This puzzle can be elucidated with a consequence of the $p$-$f$ mixing \cite{Takahashi_jpc1985,Mignod_jmmm1985}, in which the Sb 5$p$ states hybridize only with the $f\Gamma_{8}$ quartet due to having the same symmetry.
In fact, no kink structures are observed in the electron band from the Ce 5$d$ orbitals (Supplementary Note~5).
All of these results suggest the significant importance of the $p$-$f$ mixing for the multipole polaron formation.
Note that the $p$-$f$ mixing in CeSb cannot result in the Kondo screening because its low electron density of states \cite{SuzukiPhysicaB1995} is not sufficient to screen the 4$f$ moment.
The full 4$f$ magnetic moment of $\rm{Ce}^{\rm{3+}}$ ions stabilized below $T_{\rm{N}}$ strongly supports this situation.
Therefore, the discovered multipole polaron differs largely from heavy 4$f$ fermions \cite{RevModPhys.81.1551} and displays a different consequence of the significant interplay between the itinerant carriers and the localized 4$f$ states (Supplementary Note~6).

Kasuya $et$ $al$ \cite{Kasuya_lowCarrier1995} previously discussed the layered $f\Gamma_{80}$ ordering in the magnetostructures by an effective interaction of the Sb 5$p$ electrons bounded in the $f\Gamma_{80}$ layer by the magnetic $f\Gamma_{80}$-to-$f\Gamma_{7}$ excitation, called a magnetic polaron.
However, revealing its electronic correlation has remained out of reach, despite prolonged investigations.
With our direct observations, we have revealed that such a magnetic excitation ($f\Gamma_{80}$ to $f\Gamma_{7}$) does not participate in the dynamical motion of the itinerant electrons, and instead, their motions are strongly coupled to the quadrupole CEF excitation ($f\Gamma_{80}$ to $f\Gamma_{8}^{*}$).
Therefore, we can deliver our perspective of this selective electron-boson coupling to the $f\Gamma_{80}$ ordering in the devil's staircase.

\begin{figure*}
	\centering
	\includegraphics[width=\textwidth]{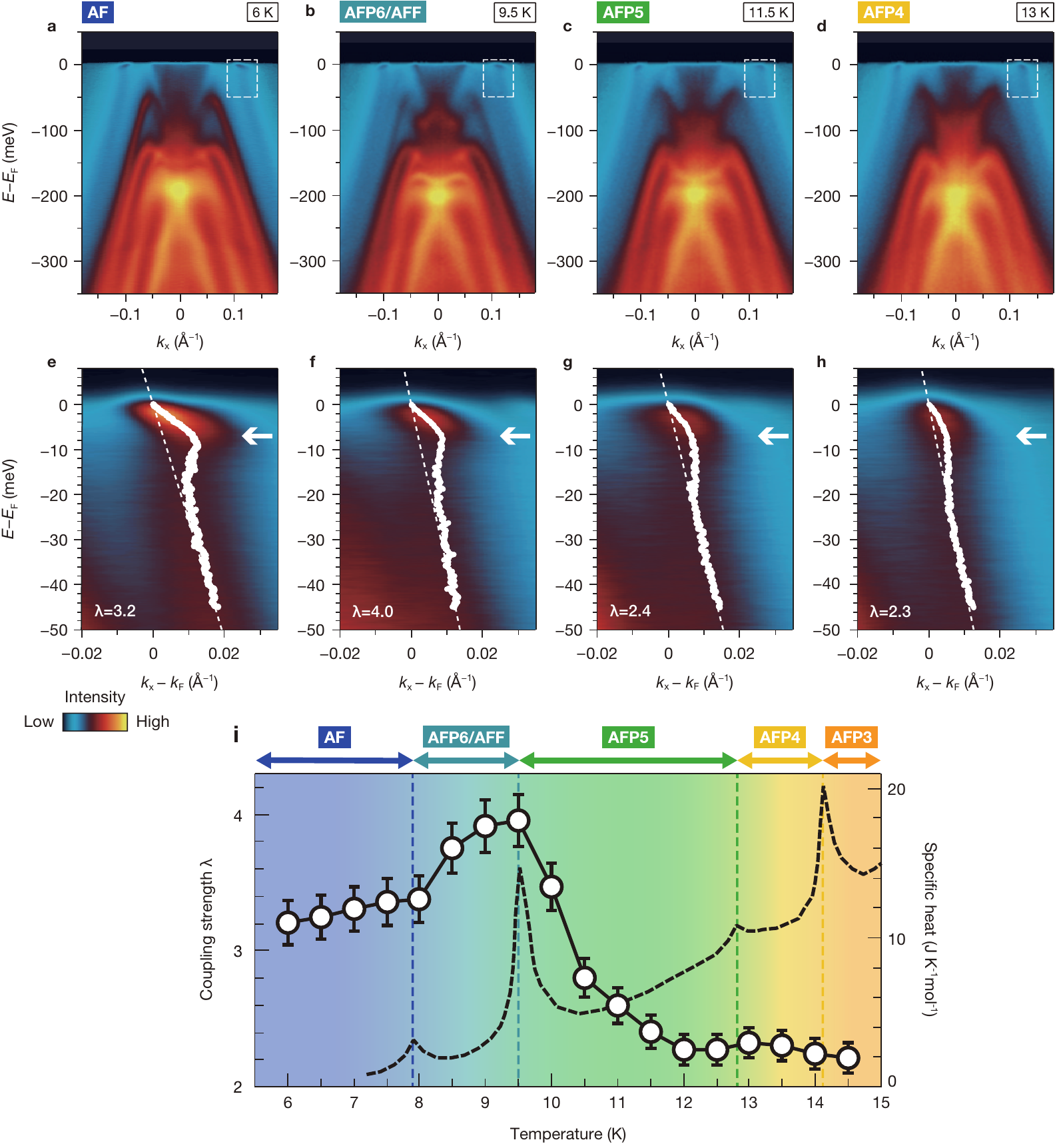}
	\caption{\label{fig4}
	\textbf{Multipole polaron throughout the devil's staircase.}
	(\textbf{a}-\textbf{d}) Representative laser ARPES maps showing different electronic reconstructions for several distinct phases \cite{kuroda2020NatureComm}.
	(\textbf{e}-\textbf{h}) Magnified maps within the narrow $E$-$k_x$ windows denoted by the rectangles in \textbf{a}-\textbf{d}, together with the peak plots by momentum distribution curve fitting (white circles).
	White arrows and dashed lines, respectively, show the kink energy and the evaluated bare bands.
	(\textbf{i}) The temperature variations of the coupling strength $\lambda$.
	The error bars are estimated in accordance with the uncertainty of the self-energy analysis.
	Background colour and dashed vertical lines illustrate each magnetic phase previously determined by specific heat measurement \cite{Mignod_jpc1980} and spectral evolution of laser ARPES \cite{kuroda2020NatureComm}.
	The black dashed line shows the specific heat curve (reproduced from Ref. \cite{Mignod_jpc1980}).
	The abrupt enhancement of the multipole polaron at the transition from the AFP5 to AFP6/AFF phase coincides with the strong jump of the specific heat.
}
\end{figure*}

To show the importance of this perspective, we focus on the temperature evolution of the coupling through the devil's staircase.
Compared to the kink structures at $20\mathchar`-200\;\rm{meV}$ observed in the other materials \cite{Valla_prl1999,Lanzara_Nature2001,Schafer_prl2004,Dahm_NaturePhys2009}, the precise measurement of the low-energy kink at $7\;\rm{meV}$ in CeSb is rather difficult, but our data obtained by laser ARPES precisely unveil the surprising evolution through the devil's staircase, as presented in Fig. \ref{fig4}.
Clear QP peaks with prominent kinks are observed not only in the AF phase (Fig. \ref{fig4}\textbf{e}) but also in the various AFP phases (Fig. \ref{fig4}\textbf{f-h}), together with the devil's staircase transition of the overall electronic structures \cite{kuroda2020NatureComm} according to the long-period 4$f$ modulations (Fig. \ref{fig4}\textbf{a-d}).
The variation of the kink energies is negligibly small, showing good agreement with the fact that no considerable temperature evolution of the bosonic-mode energy is observed by our Raman scattering (Fig. \ref{fig3}\textbf{d}).
Thus, the multipole polaron is formed throughout the devil's staircase.

Strong variation is found in $\lambda$, which is already apparent in the renormalization of the Fermi velocity in Fig. \ref{fig4}\textbf{e-h}.
Figure \ref{fig4}\textbf{i} summarizes the $\lambda$ value as a function of temperature by a $0.5\;\rm{K}$ step.
Although the kink structure in the AFP4 phase becomes less obvious (Fig.~\ref{fig4}\textbf{h}), the coupling strength is still large ($\lambda\!\approx\!2.3$), since a visible kink at low energy occurs only when the mode coupling is very strong.
There is no considerable change of $\lambda$ for the AFP3 and AFP4 phases.
However, the mode coupling is abruptly enhanced, even up to $\lambda\!\approx\!4.0$ at the transition from the AFP5 to AFP6/AFF phase (Fig.~\ref{fig4}\textbf{f}).
This enhanced $\lambda$ value is almost two times larger than that for the AFP3 and AFP4 phases. 
Another transition of the 4$f$ order at $T_{\rm{AF}}$ occurs with a slight decrease of $\lambda$ down to $\approx\!3.2$ (Fig.~\ref{fig4}\textbf{e}).
Furthermore, the changes of $\lambda$ occur sharply in the very narrow temperature range, revealing the transition-like feature.
This temperature evolution is surprising and even similar to those of the kink structure seen across the superconducting transition in cuprates \cite{Li_NaCom2018, Cuk_prl2004}.

In overall phases, the electron-boson coupling is strong ($\lambda\!\approx\!2.3\mathchar`-4.0$), which likely results from three important conditions.
One is that the low carrier density of CeSb emphasizes the role of many-body interactions~\cite{Kasuya_lowCarrier1995, SuzukiPhysicaB1995}.
As one good example, the electron-phonon coupling tends to be enhanced in the low carrier density limit because of its weak screening.
Thus, the low-carrier-density nature of CeSb likely leads to the strong mode coupling with the CEF excitation (Supplementary Note~6).
The second condition is the well-localized nature of the 4$f$ orbital states because the large density of states of the 4$f$ states sufficiently influences the electron motion.
The third condition is that the internal electronic coupling between the localized 4$f$ and Sb 5$p$ states directly impacts the QP formation, in stark contrast to the electron-phonon coupling.

It is important to find a relationship between the evolution of the multipole polaron and the devil's staircase.
The AFP magnetostructure is a consequence of the competing CEF ground states (Fig.~\ref{fig1}\textbf{b}): the ferromagnetic $f\Gamma_{80}$ plane and the paramagnetic $f\Gamma_{7}$ plane~\cite{Mignod_jmmm1985, Kasuya_lowCarrier1995}.
The presence of such competing states can often lead to a spatial modulation of the electronic interaction \cite{FobesNaturePhys2018}.
Indeed the modulation of the $p$-$f$ mixing strength was previously considered in CeSb as an origin for the AFP superstructure, where the $p$-$f$ mixing strength is periodically suppressed, leaving the paramagnetic $f\Gamma_{7}$ planes (Fig.~\ref{fig1}\textbf{b}) \cite{Iwasa_prl2002}.
The observation of no considerable change of $\lambda$ at the transitions from the AFP3 to AFP5 phase (Fig.~\ref{fig4}\textbf{i}) indicates that the multipole polaron is insensitive to the change of the $f\Gamma_{7}$ superstructures.
By sharp contrast, an abrupt enhancement of the multipole polaron occurs at the transition from the AFP5 to AFP6/AFF phase.
Notably, this enhancement coincides with the strong specific heat jump reflecting the appearance of the AFF phase, where all paramagnetic layers disappear in the whole crystal (the black dashed line in Fig. \ref{fig4}\textbf{i}) \cite{Mignod_jpc1980}.
If the AFF magnetostructure is considered, the enhancement of $\lambda$ likely occurs when all the paramagnetic $f\Gamma_{7}$ planes are eliminated in the crystal and only the $f\Gamma_{80}$ superstructure ($q\!=\!6/11$) is left (Fig. \ref{fig1}\textbf{b}).
Furthermore, the large $\lambda$ changes when the $f\Gamma_{80}$ superstructure of the AFF phase ($q\!=\!6/11$) is transformed into the simple double-layer modulation ($q\!=\!1/2$) of the AF phase.
The unusual temperature evolution of $\lambda$ sensitively varying with the elimination of the $f\Gamma_{7}$ planes and the modulation of the $f\Gamma_{80}$ superstructure is consistent with our picture of the multipole polaron that is strongly coupled to the $f\Gamma_{80}$-to-$f\Gamma_{8}^{*}$ excitation under ferromagnetic $f\Gamma_{80}$ ordering.
Also, such a correlation between the multipole polaron and the $f\Gamma_{80}$ ordering is likely related to the pseudo-gap-like behavior previously observed at the temperatures corresponding to the AFP6/AFF phase \cite{kuroda2020NatureComm}.
In addition, a sharp reduction of the resistivity across each transition was previously observed \cite{Ye_prb2018}; the conductivity becomes better with increasing (reducing) the number of the $f\Gamma_{80}$ ($f\Gamma_{7}$) planes.
Thus, our observation of the coherent QP state developing in the $f\Gamma_{80}$ ordering corresponds to these resistivity changes.

In conclusion, our systematic investigation with various probes reveals the band renormalization via the coupling of the mobile electron with the CEF excitation, and multipole polaron formation in CeSb.
The multipole polaron in CeSb exhibits extremely strong coupling and dramatically varies with ferromagnetic $f\Gamma_{80}$ ordering. 
This significant interplay between the 4$f$ order and itinerant carriers is consistent with the orbital-induced magnetoresistance anisotropy recently reported \cite{Ye_prb2018,Xu_nacom2019}, and thus our result demonstrates not only the importance of the multipole polaron picture for solving the devil's staircase puzzle, but also its high functionality for electronic device application.
The discovered coupling is possiblely due to the low-energy character of the CEF excitation, and thus this type of polaron is a unique many-body state in the 4$f$/5$f$ electron series.
Moreover, since CeSb has a very simple electronic structure \cite{Kuroda_prl2018}, this material could be regarded as a model system to further examine the fascinating correlation features between the 4$f$ order and itinerant electrons.

%%%%%%%%%%
% Figs
%%%%%%%%%%
\newpage

\newpage

\newpage

\newpage

%%%%%%%%%%
% Refs1
%%%%%%%%%%

%%%%%%%%%%
% Other info1
%%%%%%%%%%
\noindent\\
{\textbf{Methods}\\}
{\textbf{Sample growth.}\\}
CeSb single crystals were grown by the Bridgman method with a sealed tungsten crucible and high-frequency induction furnace.
High-purity Ce (99.999\%) and Sb (99.999\%) metals with the respective composition ratio were used as starting materials.
The obtained samples were characterized by the DebyeScherrer method.
\\
\noindent\\
{\textbf{Polarized microscopy measurements.}\\}
The polarized microscopy measurement was performed at the Institute for Solid State Physics at the University of Tokyo~\cite{Katakura_rsi2010}.
The sample temperature was controlled in the range of $8\mathchar`-20\;\rm{K}$.
The details of the set-up are described in our previous work~\cite{kuroda2020NatureComm}.
\\
\noindent\\
{\textbf{ARPES experiments.}\\}
The high-resolution laser ARPES with low-energy $6.994\;\rm{eV}$ ($7\;\rm{eV}$) photons was performed at the Institute for Solid State Physics at the University of Tokyo~\cite{ShimojimaJPSJ2015}.
The energy (angular) resolution was set to $2\;\rm{meV}$ ($0.3^{\circ}$).
The temperature stability during each scan was better than $\pm0.1\;\rm{K}$.
A Scienta-Omicron R4000 hemispherical electron analyzer was used, with a vertical entrance slit and light incident in the horizontal plane.
Linear polarization with the electric field aligned on the incident plane was used. 
The base pressure in the chamber was better than $1\!\times\!10^{-10}\;\rm{mbar}$.
The crystals were cleaved $in\; situ$ at the measurement temperature ($T\!=\!10\;\rm{K}$).
For the measurements of the temperature dependence, the ARPES images were scanned from lower to higher temperatures.
\\
\noindent\\
{\textbf{Raman scattering spectroscopy.}\\}
Raman scattering measurements were performed under the pseudo-backscattering configuration at $40^{\circ}$ angle of incidence.
We used an Ar laser line ($\rm{wavelength}\!=\!514.5\;\rm{nm}$) and a Jobin Yvon T64000 triple grating spectrometer (Horiba) with the spectral resolution of $\sim\!2.5\;\rm{cm^{-1}}$ corresponding to $\sim\!0.3\;\rm{meV}$.
The sample temperature was controlled in the range of $5\mathchar`-20\;\rm{K}$.
The laser power was tuned at $2\;\rm{mW}$ per spot size of $\sim\!100\;\rm{\rm{\mu m}}$ to suppress the sample overheating.
The effect of the laser heating was estimated to be less than $5\;\rm{K}$.
The sample was cleaved in the atmosphere and immediately installed in the vacuum chamber within $10\;\rm{min}$.
All spectra were corrected for the Bose-Einstein factor.
\\
\noindent\\
{\textbf{INS spectroscopy.}\\}
INS experiments were carried out to investigate the CEF and phonon excitation spectra using single-crystalline samples (Supplementary Note 4).
We used the triple-axis thermal neutron spectrometers TOPAN (6G) installed at the JRR-3 reactor in Japan Atomic Energy Research Institute, Tokai and 1T at the Orph{\'{e}}e reactor in Laboratoire L{\'{e}}on Brillouin, Saclay.
For both measurements, we used a scattered neutron energy fixed at 14.7~meV, and the sample temperatures were controlled using a helium cryostat.
\\
\noindent\\
{\textbf{Calculations.}\\}
We calculate the electron-phonon coupling of LaSb (Supplementary Note 2).
The actual calculations were carried out as follows: first, we performed an electronic structure calculation based on density functional theory with Quantum ESPRESSO code \cite{Giannozzi2017, Giannozzi_2009}.
The lattice constant of LaSb was taken from experiment \cite{Samsonov} and set to $a\!=\!6.499\;\rm{\AA}$.
We employed the fully relativistic version of Optimized Norm-Conserving Vanderbilt pseudopotentials \cite{Hamann_prb2013, Scherpelz_jctc2016}, the exchange-correlation functional proposed by Perdew {\it et al} \cite{Perdew_prl1996} and $15\!\times\!15\!\times\!15$ $k$-mesh for the self-consistent calculation.
The cut-off energy of the plane waves was set to $60\;\rm{Ry}$.
After the electronic structure calculation, we performed phonon frequency and electron-phonon vertex calculations based on density functional perturbation theory \cite{baroni_rmp2001} by using the Phonon code in Quantum ESPRESSO.
A $4\!\times\!4\!\times\!4$ $q$-mesh for the phonon momentum and $30\!\times\!30\!\times\!30$ $k$-mesh for the electron momentum were employed in the electron-phonon vertex calculation.
Finally, we calculated the Eliashberg function $\alpha^2F(\omega)$, where $\alpha$ is electron-phonon coupling strength and $F(\omega)$ is energy dependent phonon density of states, followed by the electron-phonon coupling constant $\lambda$, with the broadening width of $0.04\;\rm{Ry}$.
\\
\noindent\\
\textbf{Data availability}\\
The data that support the findings of this study are available from the corresponding author upon request.

%%%%%%%%%%
% Refs2
%%%%%%%%%%

%%%%%%%%%%
% Other info2
%%%%%%%%%%
\noindent\\
{\textbf{Acknowledgements}}\\
We acknowledge H. Kusunose, M. Kawamura and S. Tsutsui for fruitful discussions, and M. Kohgi, J.-M. Mignot and M. Braden for supports of the INS experiments.
This work was supported by Japan's Ministry of Education, Culture, Sports, Science and Technology's Quantum Leap Flagship Program (MEXT Q-LEAP, grant number JPMXS0118068681), the Ministry of Education, Culture, Sports, Science and Technology's `Program for Promoting Researches on the Supercomputer Fugaku' (Basic Science for Emergence and Functionality in Quantum Matter Innovative Strongly Correlated Electron Science by Integration of `Fugaku' and Frontier Experiments, project number hp200132), JST ERATO-FS grant number JPMJER2105, the Murata Science Foundation, Grants-in-Aid for Scientific Research (grant numbers JP21H04439, JP20H01848, JP19H00651, JP19H02683, JP19F19030, JP18H01165, JP18H01182A and JP16H06345) and Grants-in-Aids for Scientific Research on Innovative Areas, `Quantum Liquid Crystals' (grant number 19H05825) from the Ministry of Education, Culture, Sports, Science, and Technology of Japan.
\\
\noindent\\
\textbf{Author Contributions}\\
Y.A. and K. Kuroda performed the laser ARPES experiments and analysed the data.
S. Sakuragi, C.B., S.A., K. Kurokawa, S. Shin and T.K. supported the laser ARPES experiments.
Y.A., K. Kuroda, S. Sakuragi, W.-L.Z, Z.H.T, S.M. and S.T. conducted the Raman spectroscopy and analysed the data.
H.S.S., H.K. and Y.H. made the high-quality CeSb single crystals.
Y.A., K. Kuroda., S. Sakuragi, Y.K. and M.T. performed the polarizing microscopy.
K.I. performed neutron scattering spectroscopy.
T.N. and R.A. provided theoretical insights.
Y.A., K .Kuroda and T.K. wrote the paper.
All authors discussed the results and commented on the manuscript.
\\
\noindent\\
\textbf{Competing interests}\\
The authors declare no competing interests.

\end{document}